\documentclass[3p,times,procedia]{elsarticle}
\usepackage{nupha_ecrc}

\volume{00}

\firstpage{1}

\journalname{Nuclear Physics A}

\runauth{}

\jid{nupha}

\jnltitlelogo{Nuclear Physics A}

\usepackage{amssymb}
\usepackage{upgreek}

\usepackage[figuresright]{rotating}

\newcommand{\BEC}{BEC}
\newcommand{\QCD}{QCD}

\newcommand{\proton}{p}
\newcommand{\bquark}{b}
\newcommand{\cquark}{c}
\newcommand{\bbarquark}{$\mathrm{\bar{\bquark}}$}
\newcommand{\jpsi}{$\mathrm{J/\uppsi}$}
\newcommand{\mumu}{$\mathrm{\upmu^+\upmu^-}$}

\newcommand{\pp}{\proton\proton}
\newcommand{\bbbar}{\bquark\bbarquark}

\newcommand{\jpsipt}{$p\mathrm{_T^{J/\uppsi}}$}

\newcommand{\dphi}{$|\Delta \upphi^*|$}
\newcommand{\deta}{$|\Delta \upeta^*|$}
\newcommand{\ptasymmetry}{$A_\mathrm{T}$}
\newcommand{\mjpsijpsi}{$m^{\mathrm{J/\uppsi J/\uppsi}}$}
\newcommand{\ptjpsijpsi}{$p\mathrm{_T^{J/\uppsi J/\uppsi}}$}
\newcommand{\yjpsijpsi}{$y^{\mathrm{J/\uppsi J/\uppsi}}$}
\newcommand{\ptasymmetrydef}{$\left| \frac{p\mathrm{_T^{J/\uppsi_1}} - p\mathrm{_T^{J/\uppsi_2}}}{p\mathrm{_T^{J/\uppsi_1}} + p\mathrm{_T^{J/\uppsi_2}}} \right| $}

\newcommand{\cconstant}{\textit{c}}

\newcommand{\LHCbacceptance}{$2.0 < \upeta < 5.0$}

\begin{document}

\begin{frontmatter}



\title{Bose-Einstein correlations and \bbbar{} correlations in \pp{} collisions with LHCb}
\author{Bartosz Malecki on behalf of the LHCb Collaboration}
\ead{Bartosz.Malecki@ifj.edu.pl}

\address{Institute of Nuclear Physics Polish Academy of Sciences, PL-31342 Krakow, Poland}

\dochead{XXVIIth International Conference on Ultrarelativistic Nucleus-Nucleus Collisions\\ (Quark Matter 2018)}



\author{}

\address{}

\begin{abstract}
Bose-Einstein correlations for same-sign charged pions and kinematic \bbbar{} correlations in proton-proton collisions at a~center-of-mass energy of 7 and 8~TeV  are studied by the LHCb experiment. The dependence of Bose-Einstein correlation parameters on the charged-particle multiplicity is investigated. The correlation radius is observed to increase with multiplicity, while the chaoticity parameter decreases. The \bbbar{} correlations are studied using inclusive \bquark{} decays to \jpsi{} and are found to be in good agreement with theoretical predictions.
\end{abstract}


\end{frontmatter}


\section{Introduction}
\label{sec:introduction}
Studying two-particle correlations is a useful tool to get insight into the complex process of multiparticle production. Results from two different analyses of this kind are described. The first measurement concerns Bose-Einstein correlations (\BEC{}) for same-sign pions, while the second one covers kinematic \bbbar{} correlations. Both analyses are performed for data samples from high-energy proton-proton (\pp{}) collisions collected by the LHCb experiment \cite{LHCb_1,LHCb_2}. The LHCb detector is a single-arm spectrometer which covers a~pseudorapidity range of \LHCbacceptance{}. Such an acceptance in the forward region is unique among the LHC experiments and provides an additional input for understanding the process of particle production.

The \BEC{} effect emerges due to symmetrization of the total wave function of a bosonic system \cite{BEC:teo:HBT}. It can be observed as an enhancement in production of identical bosons that are close in phase space. By studying this kind of correlations a correlation radius \cite{BEC:teo:parametrization} (often interpreted as a size of a spherical static particle source at the kinetic freeze-out) and a chaoticity parameter (related to the coherence of particle emission) can be determined. Studies of the BEC effect have been performed in multiple collision systems at e.g. LEP~\cite{BEC:exp:Delphi}, RHIC \cite{BEC:exp:STAR} and LHC \cite{BEC:exp:ALICE,BEC:exp:ATLAS,BEC:exp:CMS}. Those measurements revealed many features, among them the correlation radius increasing with charged-particle multiplicity. LHCb adds a measurement in the forward direction in \pp{} collisions \cite{BEC_pp_LHCb}, which is the first result of its kind.

Heavy-flavor production provides a perfect opportunity to test quantum chromodynamics (\QCD{}). Due to their high masses, the \bquark{} and \cquark{} quarks are produced mainly in hard processes at the initial stage, which can be described in terms of perturbative \QCD{} \cite{bb:teo:general}. Studying kinematic correlations between heavy quark and antiquark can allow to identify contributions from the specific production subprocesses (e.g. gluon-splitting, flavor-excitation) and determine the size of higher-order \QCD{} corrections. The \bbbar{} correlations have been studied at e.g. SPS \cite{bb:exp:SPS}, Tevatron \cite{bb:exp:Tevatron_CDF} and LHC \cite{bb:exp:LHC_CMS,bb:exp:LHC_ATLAS}. The LHCb analysis \cite{bb_LHCb} provides first results for high-energy \pp{} collisions in the forward region, with a detector dedicated to heavy-flavor physics. The correlations are compared to predictions from PYTHIA \cite{PYTHIA} and POWHEG \cite{POWHEG} with leading-order (LO) and next-to-leading-order (NLO) calculations, respectively.

\section{Bose-Einstein correlations}
\label{sec:BEC}
The data sample used in this analysis \cite{BEC_pp_LHCb} contains $4\cdot10^7$ minimum bias events collected by LHCb in \pp{} collisions at a centre-of-mass energy $\sqrt{s}=7~\mathrm{TeV}$ . A corresponding Monte Carlo event generator simulation (MC) sample is produced using PYTHIA 8 \cite{PYTHIA}. The \BEC{} effect depends on charged-particle multiplicity $N_{ch}$, so the data is divided into three activity classes. They are defined based on a distribution of vertex locator (VELO) track multiplicity, which is a good measure of $N_{ch}$. An unfolding procedure for the VELO track multiplicities is performed using the simulation, in order to obtain the corresponding $N_{ch}$ (see~Table~\ref{tab:BEC_results}). 

Bose-Einstein correlations are studied using a correlation function. Experimentally, it is defined as \cite{BEC:teo:corr_function}:
\begin{equation}
\label{eq:correlation_function}
C_2(Q)=N(Q)^{LIKE}\ / \ N(Q)^{REF},
\end{equation}
where $Q=\sqrt{-(q_1-q_2)^2}$ is a difference of the particles four-momenta, $N(Q)^{LIKE}$ is a $Q$ distribution for pairs of signal particles and $N(Q)^{REF}$ is a $Q$ distribution for a reference sample. The signal pairs are defined as two same-sign pions coming from a single primary vertex (PV). Reference samples can be constructed in different ways, under condition that they do not contain the \BEC{} effect. In this analysis, an event-mixed reference sample is used (pairs of pions from different events). Such a solution removes the \BEC{} effect by definition, however also other types of correlations (e.g. long-range ones \cite{BEC:teo:long_range}) are not captured. Imperfections of the reference sample are reduced by introducing a double ratio, which is a ratio of $C_2(Q)$ for data to $C_2(Q)$ for MC with the BEC effect switched off. In this way, effects which are properly simulated are removed in the double ratio, such that in an ideal case it should only contain a pure BEC signal. Coulomb interactions between final-state pions are not simulated and this is corrected for by applying a Gamov penetration factor for data \cite{BEC:teo:Gamov}. The correlation function can be parametrized as \cite{BEC:teo:parametrization}:
\begin{equation}
\label{eq:correlation_function_parametrization}
C_2(Q)=N(1+\lambda \exp(-RQ))\cdot(1+\delta Q),
\end{equation}
where $R$ is the correlation radius, $\lambda$ is the chaoticity parameter, $\delta$ corresponds to long-range correlations and $N$ is a normalization factor. 

For each activity class, the double ratio is constructed and a fit with parametrization (\ref{eq:correlation_function_parametrization}) is done. 
Fit results are summarized in Table \ref{tab:BEC_results}. The systematic uncertainty (about 10\% in each activity class) is dominated by MC generator tunings and pile-up effects. The $R$ parameter increases, while $\lambda$ decreases with charged-particle multiplicity, which is consistent with previous observations on LEP and LHC. Results from the LHCb experiment are directly compared to those obtained by ATLAS \cite{BEC:exp:ATLAS}, by relating the $N_{ch}$ between both experiments acceptances using the simulation. It is observed that the LHCb results in the forward direction on both $R$ and $\lambda$ are slightly below those from ATLAS at central rapidity. 


\begin{table}
	\label{tab:BEC_results}
	\caption{Results of fits with parametrization (\ref{eq:correlation_function_parametrization}) to the double ratio in all activity classes and corresponding $N_{ch}$ ranges. Statistical and systematic uncertainties are shown (in this order). The Table is taken from \cite{BEC_pp_LHCb}.}
	\centering
	\begin{tabular}{lcccc}
		\hline
		Activity & $N_{ch}$ & $R$ [fm] & $\lambda$ & $\delta$ [GeV$^{-1}$]\\
		\hline
		Low    & [8,18]   & 1.01 $\pm$ 0.01 $\pm$ 0.10   & 0.72 $\pm$ 0.01 $\pm$ 0.05   & 0.089 $\pm$ 0.002 $\pm$ 0.044\\
		Medium & [19,35]  & 1.48 $\pm$ 0.02 $\pm$ 0.17   & 0.63 $\pm$ 0.01 $\pm$ 0.05   & 0.049 $\pm$ 0.001 $\pm$ 0.009\\
		High   & [36,96]  & 1.80 $\pm$ 0.03 $\pm$ 0.16   & 0.57 $\pm$ 0.01 $\pm$ 0.03   & 0.026 $\pm$ 0.001 $\pm$ 0.010\\
		\hline
		
	\end{tabular}
	
\end{table}

\section{Kinematic \bbbar{} correlations}
\label{sec:bb}
The \bbbar{} correlations analysis \cite{bb_LHCb} is performed using a data sample collected by the LHCb experiment in \pp{} collisions at $\sqrt{s}=7~\mathrm{and}~ 8~\mathrm{TeV}$. The beauty hadrons are reconstructed from their inclusive decays to \jpsi{} mesons in a kinematic range $2.0 < y^{\mathrm{J/\uppsi}} < 4.5$, $2 < p\mathrm{_T^{J/\uppsi}} < 25$~GeV/\cconstant{}, where \jpsi{} further decays to \mumu{} pair. The analysis is performed for four different requirements on \jpsi{} transverse momentum \jpsipt{}.

For each \jpsipt{} region, a two-dimensional distribution of the \mumu{} masses for the selected \jpsi{} pairs is constructed. From such distributions, the signal yield is determined by fitting a function which describes both the signal term (two \jpsi{} mesons) and background (\mumu pairs that do not originate from \jpsi{} decay). Correlations are studied for a number of variables by constructing normalized differential production cross-sections \cite{bb:teo:cross-section}:
\begin{equation}
\label{eq:cross-section}
\frac{1}{\sigma} \frac{\mathrm{d}\sigma}{\mathrm{d}v} \equiv \frac{1}{N^{cor}} \frac{\Delta N_i^{cor}}{\Delta v_i},
\end{equation}
where $v$ is a generically denoted kinematic variable, $N^{cor}$ is the total number of efficiency-corrected signal candidates and $\Delta N_i^{cor}$ is the number of efficiency-corrected signal candidates in bin $i$ of width ${\Delta v_i}$. The efficiency correction \cite{bb:teo:efficiency} is based on both simulation and data-driven methods and takes into account efficiencies related to: the LHCb detector acceptance, reconstruction and selection of \jpsi{} candidates, the muon identification and trigger. The variables used in the analysis are: difference in the azimuthal angle \dphi{} and pseudorapidity \deta{} of the two beauty hadrons\footnote{Both variables are estimated on the basis of the direction of the vector from the PV to the \jpsi{} decay vertex.}, mass \mjpsijpsi{}, transverse momentum \ptjpsijpsi{} and rapidity \yjpsijpsi{} of the \jpsi{} pair, and the asymmetry between the $p\mathrm{_T}$ of \jpsi{} mesons \ptasymmetry{}~$\equiv$~\ptasymmetrydef{}. In each case, the systematic uncertainties are much smaller than the statistical ones (most of them cancel out in the definition of normalized differential production cross-section) and they are neglected.

\begin{figure}[htb]
	\centerline{%
		\includegraphics[width=15.5cm]{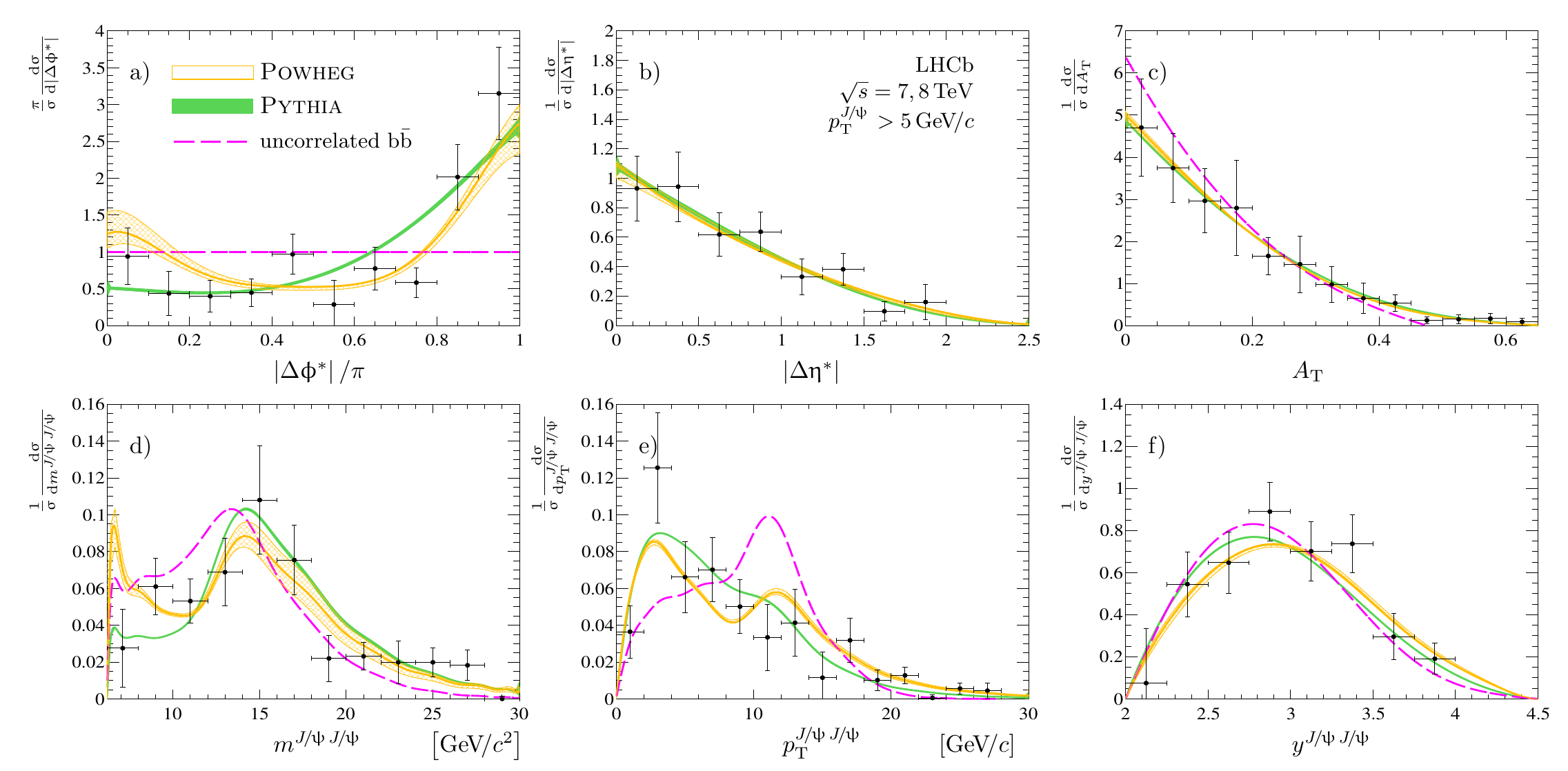}				
	}
	\caption{Normalized differential production cross-sections (points with error bars) in \jpsipt{} $>$ 5~GeV/\cconstant{} region for a) \dphi{} /$\pi$, b) \deta{}, c)~\ptasymmetry{}, d) \mjpsijpsi{}, e) \ptjpsijpsi{}, f) \yjpsijpsi{}. Expectations from POWHEG (orange line), PYTHIA (green band) and model of uncorrelated \bbbar{} production (dashed magenta line) are shown. The Figures are taken from \cite{bb_LHCb}.}
	\label{fig:bb}
\end{figure}

Results are compared with expectations from PYTHIA (LO) and POWHEG (NLO), as well as with a~data-driven model of uncorrelated \bbbar{} production (see Fig.~\ref{fig:bb}). It can be observed that both PYTHIA and POWHEG well describe the data for all distributions. This suggests that the effect of the NLO correction on the \bbbar{} production correlations is small, at least compared to the experimental precision.

\section{Summary}
\label{sec:summary}
Bose-Einstein correlations for same-sign charged pions and kinematic \bbbar{} correlations in \pp{} collisions at $\sqrt{s}=7~\mathrm{and}~ 8~\mathrm{TeV}$ are studied by the LHCb experiment. Both analyses are the first kind of this measurements in the forward region and show a good potential of LHCb in the respective fields.

The correlation radius measured in the BEC study increases with charged-particle multiplicity, while the chaoticity parameter decreases. This is consistent with other observations on LEP and LHC. The BEC parameters measured by LHCb in the forward region seem to be slightly lower than those observed by ATLAS for corresponding charged-particle multiplicities. A more detailed comparison is planned by performing a~full three-dimensional analysis at LHCb.

The \bbbar{} correlations are studied using inclusive \bquark{} decays to \jpsi{}. The observed correlations are well described by both PYTHIA and POWHEG, which suggests that the effect of NLO corrections on the correlations is small compared to the experimental precision. The differences between PYTHIA and POWHEG calculations are larger for higher \jpsipt{} regions. Discriminating between theory predictions is not possible with the present data, however future analyses will profit from larger samples.
\newline\newline\textbf{Acknowledgments}\newline
This work was supported by PL-GRID and by NCN (Poland) under the contract no. 2013/11/B/ST2/03829.





\bibliographystyle{elsarticle-num}


\end{document}